\documentclass{aastex631}
\usepackage{amsmath}

\begin{document}

\title{\textsc{nifty-ls}: Fast and Accurate Lomb-Scargle Periodograms Using a Non-Uniform FFT}

\author[0000-0002-9853-5673]{Lehman H. Garrison}
\affiliation{Scientific Computing Core,
Flatiron Institute,
162 Fifth Avenue,
New York, NY 10010, USA}

\author[0000-0002-9328-5652]{Dan Foreman-Mackey}
\affiliation{Center for Computational Astrophysics,
Flatiron Institute,
162 Fifth Avenue,
New York, NY 10010, USA}

\author{Yu-hsuan Shih}
\affiliation{NVIDIA Corporation,
2788 San Tomas Expy,
Santa Clara, CA 95051, USA}

\author{Alex Barnett}
\affiliation{Center for Computational Mathematics,
Flatiron Institute,
162 Fifth Avenue,
New York, NY 10010, USA}



\begin{abstract}
We present \textsc{nifty-ls}, a software package for fast and accurate evaluation of the Lomb-Scargle periodogram. \textsc{nifty-ls} leverages the fact that Lomb-Scargle can be computed using a non-uniform FFT (NUFFT), which we evaluate with the Flatiron Institute NUFFT package (\textsc{finufft}). This approach achieves a many-fold speedup over the Press \& Rybicki (1989) method as implemented in \textsc{Astropy} and is simultaneously many orders of magnitude more accurate. \textsc{nifty-ls} also supports fast evaluation on GPUs via CUDA and integrates with the \textsc{Astropy} Lomb-Scargle interface. \textsc{nifty-ls} is publicly available as open-source software.
\end{abstract}



\defcitealias{1989ApJ...338..277P}{PR89}

\section{Introduction} \label{sec:intro}

The Lomb-Scargle periodogram is a widely-used analysis tool for astronomical time-series data\footnote{For recent improvements in time-series analysis with unevenly sampled data, see \cite{2024arXiv240413223W} or \cite{2024AJ....167...22D}, which also use \textsc{finufft}.} (\citealt{1976Ap&SS..39..447L,1982ApJ...263..835S}; c.f.~\citealt{2018ApJS..236...16V}). However, its direct computation is prohibitively slow: for a number of observations $N_d$ and number of periodogram frequencies $N_f$, the work has time complexity $\mathcal{O}(N_dN_f)$. The popular \cite{1989ApJ...338..277P} method (hereafter \citetalias{1989ApJ...338..277P}) addresses this by posing the computation as four weighted trigonometric sums solved with a pair of fast Fourier transforms (FFTs) on an equi-spaced grid. The unevenly sampled data points are interpolated to this grid by a process that \citetalias{1989ApJ...338..277P} call ``extirpolation.''  The trigonometric sums take the form:
\begin{align}
&S_k = \sum_{j=1}^{N_d} h_j \sin(2 \pi f_k t_j),
&C_k = \sum_{j=1}^{N_d} h_j \cos(2 \pi f_k t_j),
\quad k=0,1,\dots,N_f,
\label{eqns:pr89}
\end{align}
where
$f_k$ is the cyclic frequency of bin $k$, $t_j$ are the observation times (of which there are $N_d$), and $h_j$ are the weights. The \citetalias{1989ApJ...338..277P} method
 approximates these sums via extirpolation, which, with the FFT, reduces the work to $\mathcal{O}(N_d + N_f\log N_f)$. This is the default algorithm in popular packages like \textsc{Astropy} \citep{2018AJ....156..123A} and \textsc{LombScargle.jl}\footnote{\url{https://github.com/JuliaAstro/LombScargle.jl}}. Indeed, the algorithm is simply called \texttt{fast} by \textsc{Astropy}\footnote{\url{https://docs.astropy.org/en/v6.1.x/timeseries/lombscargle.html}}.

We observe \citep[following, e.g.,][]{2012A&A...545A..50L} that Eqs.~\ref{eqns:pr89} are exactly the equations of a non-uniform fast Fourier transform (NUFFT). A NUFFT is an extension of the FFT to unevenly sampled data \citep{dutt}.  Specifically, a type 1 (non-uniform to uniform, also known as ``adjoint'') complex NUFFT evaluates
\begin{equation}
g_k = \sum_{j=1}^{N_d} h_j e^{i k t_j},
\qquad k=-N_f/2, -N_f/2+1, \dots, N_f/2 .
\end{equation}
In modern NUFFT algorithms, the accuracy of evaluation $\varepsilon$ is a user-controlled parameter that may be chosen as small as desired, down to nearly machine precision. Higher accuracy increases the cost per nonuniform data point:
specifically, the NUFFT time cost is $\mathcal{O}(N_d \log \varepsilon^{-1} + N_f \log N_f)$.
We see that the real and imaginary parts of $g_k$ are \citetalias{1989ApJ...338..277P}'s $C_k$ and $S_k$ respectively, for frequencies $f_k = k/2\pi$.



We evaluate the NUFFT with \textsc{finufft}, the Flatiron Institute NUFFT \citep{2019SJSC...41C.479B,BARNETT20211}.  The \textsc{finufft} approach is similar in spirit to \citetalias{1989ApJ...338..277P}---upsample the data to a fine grid, perform an FFT, then correct with a diagonal rescaling---but \textsc{finufft} has a near-optimal spreading kernel (an ``exponential of semicircle'') that it uses instead of extirpolation. Its width and fine grid size are carefully chosen to match the user tolerance $\varepsilon$, and manual SIMD-vectorized piecewise polynomials makes kernel evaluation extremely fast. \textsc{finufft} also supports multi-threaded CPU evaluation and GPU evaluation via CUDA \citep{9460591}, with best-in-class performance and accuracy.


We implement this \textsc{finufft}-based Lomb-Scargle periodogram in the \textsc{nifty-ls} software package. \textsc{nifty-ls} is written in Python, with some modules implemented using C\texttt{++} and OpenMP for performance. It provides a backend to the \textsc{Astropy} Lomb-Scargle interface, so code that uses \textsc{Astropy} can benefit with minimal changes (setting \texttt{method="fastnifty"} in \textsc{Astropy} calls). \textsc{nifty-ls} is developed as open-source software on GitHub\footnote{\url{https://github.com/flatironinstitute/nifty-ls/}}, with binary distributions available on PyPI.



\section{Results}

\begin{figure}
\plottwo{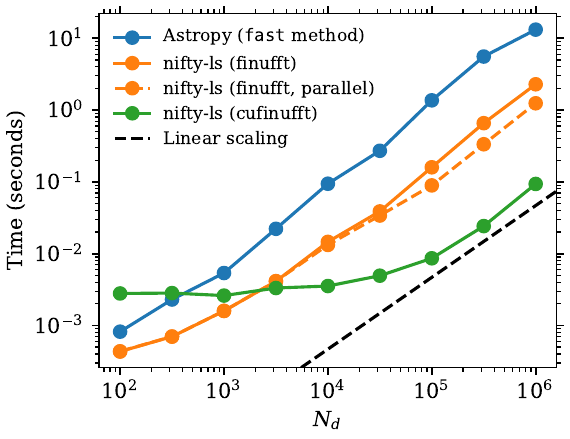}{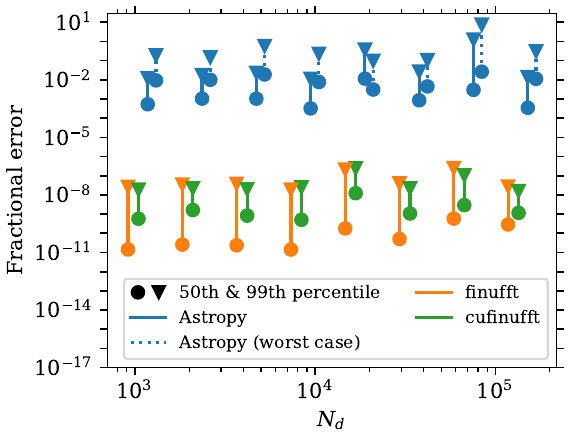}
\caption{\textit{Left panel:} the time to compute the periodogram of a time series with $N_d$ observations and $N_f \approx 12 N_d$ frequencies. The ``\textsc{Astropy} \texttt{fast}'' method is the \citetalias{1989ApJ...338..277P} method. The three \textsc{nifty-ls} methods use the \textsc{finufft} (CPU), \textsc{finufft} (multi-threaded CPU), and \textsc{cufinufft} (CUDA GPU) backends. \textit{Right panel:} the accuracy of the periodogram. The ``\textsc{Astropy} (worst case)'' and ``\textsc{Astropy}'' results use a number of frequency bins just before and after the threshold that yields a factor-of-2 jump in the size of the internal FFT, typically increasing accuracy. The plotted $N_d$ values are staggered for clarity.
\label{fig:results}}
\end{figure}

We characterize the performance and accuracy of \textsc{nifty-ls} in Fig.~\ref{fig:results}.  We use \textsc{nifty-ls} version 1.0.1, \textsc{finufft} version 2.3, and \textsc{Astropy} version 6.1.3.  As a test dataset, we use a double-precision time series with $N_d \in \left[10^2, 10^6\right]$ and use the default \textsc{Astropy} $N_f \approx 12 N_d$.  We use (up to) 16 cores of a Intel Xeon Ice Lake processor for the \textsc{finufft} (CPU) backend and and 1 NVIDIA A100 GPU for the \textsc{cufinufft} (GPU) backend.

\subsection{Performance}
\textsc{nifty-ls} is several times faster than \textsc{Astropy} \texttt{fast} on using a single CPU thread, with speedups of $2\times$ on small transforms to $5\times$ on large. This is seen in the left panel of Fig.~\ref{fig:results}. CPU multi-threading helps large transforms, yielding a $11\times$ speedup over \textsc{Astropy}, although the improvement is sub-linear in the number of threads (a well-known result for FFTs, since the time is dominated by memory accesses). 
The performance can be summarized by around $10^7$ output frequencies per second on the CPU, or $10^8$ on the GPU. Additionally, \textsc{nifty-ls} exposes a batched processing mode for sets of time series with the same observation times, which can more efficiently use CPU and GPU parallelism.

Using the GPU via \textsc{cufinufft}, \textsc{nifty-ls} is $200\times$ faster than \textsc{Astropy} for large $N_d$. For small $N_d$, the problem is not large enough to overcome GPU overheads or saturate the GPU. Still, for $N_d$ as small as 4000 there is a benefit over \textsc{finufft}, which extends to smaller $N_d$ for batched transforms.

Our tests also suggest that the NUFFT approach is more efficient than a brute-force (i.e.~quadratic complexity) CUDA evaluation of Eqs.~\ref{eqns:pr89} in the style of \cite{Townsend_2010} and \cite{2021A&C....3600472G}. The latter work reports a single-object rate of about $2\times 10^7$ frequencies per second on a P100 GPU, which is about the speed of our CPU version, and several times slower than our GPU version even after accounting for the naive $2.4\times$ performance advantage of the A100.

\subsection{Accuracy}
\textsc{nifty-ls} is many orders of magnitude more accurate than \textsc{Astropy} \texttt{fast}, using default settings for both. On this dataset, it is conservatively 6 orders of magnitude better. This is shown in the right panel of Fig.~\ref{fig:results}. For each method, the median and 99\textsuperscript{th} percentile of the periodogram error distribution is shown.
Furthermore, for \textsc{Astropy}, two cases are considered.  Internally, \textsc{Astropy} uses an FFT whose size is the next power of 2 above the specified oversampling rate. A jump in FFT size typically yields an increase in accuracy. Thus, we expect the worst accuracy to be from the largest $N_f$ that does not yield such a jump. This is labeled ``Astropy (worst case)'' in the Figure. The case labeled ``Astropy'' is the first $N_f$ after such a jump.

For \textsc{Astropy} in the baseline case, median errors are about 0.1\%, while the 99\textsuperscript{th} percentile error is a few percent (with a few excursions to larger errors). In the worst case, however, the median error is typically 1\% and the 99\textsuperscript{th} percentile is 10\% to 100\% or greater.

\textsc{nifty-ls} demonstrates greater accuracy for both backends, with median relative errors of $10^{-11}$ to $10^{-8}$, and 99\textsuperscript{th} percentile errors of $10^{-8}$ to $10^{-6}$. We note that the condition number of the problem\footnote{See \url{https://finufft.readthedocs.io/en/latest/trouble.html}}
imples that errors in Eqs.~\ref{eqns:pr89} cannot in principle be smaller than $\mathcal{O}(N_f\varepsilon_\textrm{mach})$, where $\varepsilon_\textrm{mach}$ is machine precision. 
For double precision, $\varepsilon_\textrm{mach} \approx 10^{-16}$, so the minimum achievable error for a transform of length $N_f=10^5$ is about $10^{-11}$. Were we to use single precision, with $\varepsilon_\textrm{mach} \approx 10^{-7}$, the minimum achievable error would be quite large, about $10^{-2}$. Hence, we restrict our tests to double precision and recommend its use in large periodograms.



\textsc{Astropy} has several parameters that can be tuned to yield more accurate results, but these come at the cost of performance. Likewise, \textsc{nifty-ls} can produce lower-precision results
slightly faster than reported above, but with $N_f\approx 12N_d$, the FFT time dominates over the spreading time.


The reference result here is a direct evaluation of the trigonometric sums in Eqs.~\ref{eqns:pr89}, using exact angle-sum trigonometric identities to accelerate the computation (a ``phase winding'' method). Using \textsc{Astropy} \texttt{fast} with ``\texttt{use\_fft=False}'' (which does a brute-force evaluation) yields the same result.

In summary, \textsc{nifty-ls} provides greater accuracy and speed than extirpolation-based methods, and thanks to its \textsc{Astropy} integration and available binary distributions, has a low barrier to adoption. We hope that this will provide an improved experience for users of the Lomb-Scargle periodogram.

\begin{acknowledgments}
LG would like to thank Soichiro Hattori for early feedback, testing, and fixes.
\end{acknowledgments}






\bibliography{biblio}{}
\bibliographystyle{aasjournal}



\end{document}